\documentclass[prl,twocolumn,superscriptaddress]{revtex4}
\usepackage{graphicx}
\usepackage{epstopdf}
\usepackage{epsfig}
\usepackage{times}
\usepackage{amsmath}
\usepackage{color}
\usepackage[squaren]{SIunits}
\usepackage{hyperref}

\begin{document}

\title{Comment on ``Quantum Coherence and Sensitivity of Avian Magnetoreception''}

\author{Erik M. Gauger}
\email{erik.gauger@materials.ox.ac.uk}
\affiliation{Centre for Quantum Technologies, National University of Singapore, Singapore}
\affiliation{Department of Materials, University of Oxford, Parks Rd, Oxford OX1 3PH, UK}

\author{Simon C. Benjamin}
\affiliation{Department of Materials, University of Oxford, Parks Rd, Oxford OX1 3PH, UK}
\affiliation{Centre for Quantum Technologies, National University of Singapore, Singapore}

\begin{abstract}
In a recent Letter [Phys. Rev. Lett. \textbf{109}, 110502 (2012)], Bandyopadhyay, Paterek and Kaszlikowski report their analysis of spin coherence time in the radical pair involved in avian magnetoreception, concluding that is of the order of a microsecond. However, a combination of an erroneous numerical calculation together with an incorrect parameter drawn from an experimental source have resulted in the authors underestimating by two orders of magnitude. Consequently, one must reverse the authors' conclusion that the timescale is consistent with experiments on cryptochrome.
\end{abstract}

\maketitle

Several papers \cite{cai2010,gauger2011,cai2012} have studied the quantum physics of the radical pair (RP) mechanism hypothesised to underlie the avian compass. 
Our 2011 Letter~\cite{gauger2011} analysed the coherence time of the electron spin pair and found that it must be surprisingly long. To be consistent with behavioural studies on European Robins involving weak radio frequency (rf) fields~\cite{ritz04, ritz09}, the coherence time should be of order \unit{100}{\micro\second} or more. Interestingly this is considerably longer than the reported \unit{6}{\micro\second} radical pair lifetime from {\it in vitro} experiments on cryptochrome \cite{biskup2009}, widely considered a potential candidate for the avian compass \cite{lau2012}. 

Utilising the radical pair model we described in Ref.~\cite{gauger2011}, Bandyopadhyay, Paterek and Kaszlikowski (BPK) seek to close this gap by considering additional behavioural studies, as reported in a very recent Letter~\cite{bandyopadhyay2012}.  However, their analysis suffers from two errors: an erroneous numerical computation together with the omission of vital experimental data. These issues are multiplicative and result in an underestimate of the lower bound by a factor of about 40.  
Consequently, the estimate of the lifetime given in the paper as \unit{5-6.7}{\micro\second}, and described as ``of the order of a microsecond'' in their abstract, in fact becomes \unit{200-270}{\micro\second}, i.e., hundreds of microseconds.

To test the validity of BPK's numerical calculation, we regenerated their simulation results using exactly the model and the parameters which they select. After an exhaustive series of  simulations, we conclude that it is not possible to reproduce the graphs in BPK's Letter. One can match the line shapes exactly, but to do so one must rescale by a factor of four either the time axis or the spins' g-factors. In an online document~\cite{supplement}, we provide complete details of our analysis for scrutiny. Furthermore, we have been made aware that an independent researcher also found it impossible to reproduce BPK's results without artificially scaling the model parameters~\cite{Lambert}. Evidently, there is an error in the numerical code employed by BPK.

In deriving lifetime estimates, both our original Letter and BKP's vitally depend on the effect of weak resonant fields in disrupting the birds' compass sense.  Experimentalists have reported disruptions for fields of strength \unit{470}{\nano\tesla} to \unit{15}{\nano\tesla}. In our paper we took the value of $\unit{150}{\nano\tesla}$ to ensure a conservative estimate; however, to argue that a specific shorter process timescale is consistent with the body of behavioural experiments, the analysis should be based on the {\it weakest} rf field known to disrupt the bird's compass sense, i.e.~\unit{15}{\nano\tesla}. BPK perform their calculations for $B_{\mathrm{rf}}= 470,~150,$ and \unit{47}{\nano\tesla} but inexplicably they omit the crucial \unit{15}{\nano\tesla} datum (see Fig. 3 of Ref.~\cite{ritz09}, which BPK cite as their Ref.~[13]). The effect of including this result is to increase the lower bound on the lifetime by a factor of about $10$, which becomes $40$ in view of the numerical error described above~\cite{supplement}. Stated alternatively: the timescale reported by BKP is not consistent with the reported disruption of the avian compass at fields of \unit{15}{\nano\tesla}; any bird whose compass lifetime is confined to microseconds (or indeed 10s of microseconds) must be immune to a \unit{15}{\nano\tesla} oscillatory field. 

BKP's observation that long coherence is not {\em required} for a compass sense remains correct. However, this is not a novel observation, having been stated and analysed in our 2011 Letter~\cite{gauger2011} and in Ref.~\cite{cai2012}; the latter specifically examined the cases where noise is beneficial. Notwithstanding the puzzle of why the bird should evolve an unnecessarily long  lifetime \cite{stoneham2012},  the available data~\cite{ritz04, wiltschko2006, ritz09} applied to a proper quantum mechanical model of the RP mechanism nevertheless indeed imply that the life- and coherence time is of order \unit{100}{\micro\second} or more.

\end{document}